\documentclass[a4paper,12pt]{article}
\usepackage[utf8]{inputenc}
\usepackage{amssymb,amsmath}
\usepackage{graphicx}
\usepackage{hyperref}
\title{Entropy spectrum of BTZ black hole in massive gravity}
\author{Jishnu Suresh$^{1}$ and V. C. Kuriakose$^2$ \\
\vspace{0.2in} \\
 Department of Physics, \\
Cochin University of Science and Technology, 
\\{Kochi-22, India}.\\
\vspace{0.1in} \\
E-mail: $^1$ jishnusuresh@cusat.ac.in \\
$^2$vck@cusat.ac.in\\}

\begin{document}
\date{}
\maketitle

\begin{abstract}
We study the entropy spectrum of (2+1) BTZ black holes in massive gravity models. We use the formalism proposed by Jiang and Han where black hole property of adiabaticity and the oscillating velocity of the black hole horizon are used to investigate the quantization of the entropy of such systems. We find that the entropy of the BTZ black holes in massive gravity is quantized with equally spaced spectra. 
\end{abstract}

\section{Introduction} \label{intro}
A black hole in three dimensional space time (2+1)  can be acted as a perfect toy model to study the gravity theories as well as the black hole properties of the system. This space time is commonly known as Banados
-Teitelboim-Zanelli (BTZ) black hole \cite{BTZ1,BTZ2,BTZ3}. The salient feature of this toy model lies in its simplicity of its construction. Despite its simplicity, the BTZ black hole plays a good role in the understanding of thermodynamics of black holes \cite{BTZthermo1,BTZthermo2,BTZthermo3,BTZthermo4,BTZthermo5,BTZthermo6,BTZthermo7}, recent developments in string theory \cite{BTZstring1,BTZstring2,BTZstring3,BTZstring4}, as well as many thermodynamic studies in the presence of different fields and different gravity theories  \cite{BTZZ1,BTZZ2,hendi, BTZZ3,BTZZ4,BTZZ5,BTZZ6}. In this paper we will focus on the studies of BTZ black hole, the (2+1) toy model, in the context of massive gravity theory. 

We know that Einstein's General Relativity  describes nonlinear self interactions of massless spin 2 excitations or in other words gravitons as massless particles of two degrees of freedom. To consider a massive graviton, the GR must be modified. The most natural way of modifications would be 
by adding a mass term for the spin 2 field. As a result, the modified theory would explain the nonlinear interactions of a massive spin 2 field, and 
this theory is widely known as Massive gravity. The first attempt in this direction was happened in 1939, when Fierz and Pauli \cite{fp} considered modification by adding a mass term to a linearized theory of gravity.
The proposed theory was unique in such a way that there were only a single way to add a mass term so that the theory becomes physically significant. 
Later in two independent articles, van Dam and Veltman and Zakharov claimed that Fierz-Pauli theory could not converge to Einstein's theory
in the zero mass limit \cite{vd,zak}.  Later in 1972, Vainshtein introduced a new mechanism to overturn the above mentioned 
vDVZ discontinuity \cite{vain}. In this mechanism he considered the full non linear formulation of massive gravity and as a result in the 
zero mass limit Einstein's original equations are retrieved. In the same year, Boulware and Deser proved that any non linear massive gravity theory
which uses the Vainshtein's mechanism would contain a 'ghost' \cite{bd1,bd2}, known as the BD ghost. The presence of BD ghost remained as an unsolved 
problem until 2010, when de Rham, Gabadadze and Tolley (dRGT) proposed the first non linear completion of the FP theory 
free of BD ghost instability \cite{drgt1,drgt2}. They showed that the potential to be ghost free up to the quartic
order in perturbation and to all orders in decoupling limit, and as a result of this many extensions of this theory 
are discovered \cite{has1,has2,has3,rosen1,rosen2,aragone,isham,salam}. Besides these extensions, alternative theories with massive graviton have also  been under rigorous investigations. These theories include the DGP model \cite{dgp1,dgp2}, Kaluza - Klein scenarios \cite{kk,kkk}, New massive gravity \cite{bht} and Topological Massive gravity \cite{deser1,deser2}. Black hole solutions in the presence of such theories were obtained \cite{NewBlack1,NewBlack2,NewBlack3,NewBlack4,NewBlack5,NewBlack6,NewBlack7,dRGTb1,dRGTb2,dRGTb3}. To have a deep knowledge in the case of three dimensional massive gravity, one can relay on the electromagnetic field and consider it as the source. In this paper, we will consider the three dimensional BTZ black holes in the case of both New massive gravity and Einstein Maxwell massive gravity. In the NMG case, we considered the chargless BTZ solutions, where as in the second case an electromagnetic coupled, hence charged BTZ solution is investigated. 

After the discovery of Hawking radiation, it was widely believed that the black hole physics studies may shed light to the understandings of quantum theory of gravity. Bekesntein \cite{bekenstein} was the first to begin the investigation in this direction. As a result he concluded that the horizon area of a non-extremal black hole must have a discrete spectrum.  In order to construct the discrete area levels, Bekenstein found a lower bound for the increase in the black hole event horizon area as,
\begin{equation}
(\Delta A)_{min}= 8 \pi l_{p}^{2}
\end{equation}
where $l_{p}= (\frac{G \hbar}{c^{3}} )^{1/2}$ is the Planck length. It was interesting enough to note that the lower bound in the change in area does not depend on black hole parameters. Hence it was to consider as an evidence for equispaced area spectrum devoid of any black hole parameters.

These ideas put forward by Bekenstein became the corner stone of the investigation in the direction of the calculations and derivations of the area and hence the entropy spectrum of a black hole. Quasinormal mode (QNM) frequencies are known as the characteristic sound of the black hole. Hence these QNMs must have an adiabatic invariant quantity. As a result of the existence of such a relation, Hod \cite{hod1,hod2} derived the area as well as entropy spectrum of black hole from QNMs. Using Bohr-Sommerfield quantization rule, given as,
\begin{equation}
I_{\mbox{adiabatic}}= n \hbar,
\label{BS_rule}
\end{equation}
he found that the Schwarzschild black hole area spectrum is equispaced. From this area spectrum one can arrive at the entropy spectrum of the black hole by using the well known Bekenstein-Hawking area law as,
\begin{equation}
\Delta S_{\mbox{bh}}= \ln 3.
\end{equation}
Later Kunstatter \cite{kunstatter} calculated the area spectrum of d-dimensional spherically symmetric black holes by considering the explicit form of the adiabatic invariant quantity as,
\begin{eqnarray}
I_{\mbox{adiabatic}}= \int \frac{dE}{\Delta \omega(E)}, \\
\Delta \omega = \omega_{n+1} - \omega_n,
\end{eqnarray}
where $E$ and $\omega$ are respectively energy and frequency of the QNM. He obtained the area spectrum of the black hole by considering (\ref{BS_rule}), and that yields the same spacing obtained by Hod. In this work, Hod and Kunstatter considered the real part of the QNM frequency to calculate the area spectrum. Maggiore refined the idea proposed by Hod by providing a new interpretation \cite{maggiore} where, black hole is found to behave like a damped harmonic oscillator for which the physical frequency of QNM is determined by its real and imaginary part as,
\begin{equation}
\omega= \sqrt{\omega^2 _R+ \omega^2 _I}.
\end{equation}

Recently Majhi and Vagenas \cite{majhi} proposed a new method to quantize the entropy without using quasinormal modes. They used the idea of relating an adiabatic invariant quantity to the Hamiltonian of
the black hole, and they obtained an equally spaced entropy spectrum with its quantum to be equal to the one obtained by Bekenstein. When the tunneling picture is incorporated, the black hole horizon can be assumed to oscillate periodically when the particle tunnels in or out. Interestingly this approach follows Maggiore's method in which the  
perturbed black hole behaves as damped harmonic oscillator. In this tunneling picture, we begin with the  adiabatic invariant quantity of the form,  which is basically the action of the oscillating horizon,
\begin{equation}
I=\int p_i d q_i,
\label{action_AI}
\end{equation}
where $p_i$ is the corresponding conjugate momentum of the coordinate $q_i$ and $i=0,1$ for which $q_0=\tau$ and $q_1=r_h$. Here, $\tau$ represents the Euclidean time and $r_h$ is the horizon radius. By implementing the Hamilton's equation $\dot{q}_i= \frac{d H}{d p_i}$, where $H$ denotes the Hamiltonian of the system, which acts as total energy of the black hole, into (\ref{action_AI}), one can rewrite the action as,
\begin{equation}
I=\int p_i d q_i = \int  \int_{0} ^{H} dH^\prime d\tau 
 +\int \int_{0} ^{H} \frac{dH^\prime}{\dot r_{h}} dr_h
  =2\int  \int_{0} ^{H} \frac{dH^\prime}{\dot r_{h}} dr_h~.
\label{action_AI_modi}
\end{equation}
where $\dot r_{h} = \frac{d r_h}{d \tau}$. 
In order to calculate the above adiabatic invariant quantity, let us consider static, spherically symmetric black hole solution, in general given by,
\begin{equation}
ds^2 = -{N(r)}^2dt^2 +\frac{dr^2}{f(r)}+ r^2 d\Omega^2~.
\label{metric_general_intro}
\end{equation}
One can obtain $r_h$, namely the black hole horizon radius from the relation $N(r_h)=f(r_h)=0$. 
To evaluate the integral (\ref{action_AI_modi}), one must find the oscillating velocity of the black hole horizon. In the tunneling picture, when a particle tunnels in or out, the black hole horizon will expand or shrink due to gain and loss of the black hole mass. Since the tunneling and oscillation happens simultaneously, then the tunneling velocity of particle is equal and opposite to the oscillating velocity of the black hole horizon, i.e,
\begin{equation}
 \dot r_{h}= -\dot r ~.
 \end{equation}
In (\ref{action_AI_modi}), $\tau$ denotes the Euclidean time, hence one has to Euclideanize the metric given by (\ref{metric_general_intro}), by introducing the transformation $t \rightarrow -i \tau$. Then,
\begin{equation}
ds^2 ={N(r)}^2d \tau^2 +\frac{dr^2}{f(r)}+ r^2 d\Omega^2~.
\end{equation}
Now, when a photon travels across the black hole horizon, the radial null path $(ds^2=d\Omega^2=0)$, often called as radial null geodesic is given by, 
\begin{equation} 
\dot r=\frac{dr}{d\tau}=\pm i \sqrt{N\left(r\right)^2 f\left(r\right)}~,
\label{radial null path}
\end{equation}
where the positive sign denotes the outgoing radial null paths and negative sign represents the incoming radial null paths. Among these we will consider the outgoing paths for area spectrum calculations, because these paths are more related to the quantum behaviours under consideration. Hence the shrinking velocity of the black hole horizon is,
\begin{equation}
 \dot r_{h}= -\dot r = -i \sqrt{N\left(r\right)^2 f\left(r\right)}.
 \end{equation}
 Then, (\ref{action_AI_modi}) is  now read off,
 \begin{equation}
\int p_i d q_i   =-2i \int  \int_{0} ^{H} \frac{dH^\prime}{\sqrt{N\left(r\right)^2 f\left(r\right)}} dr~.
\end{equation}
To find this adiabatic invariant quantity, one has to execute this integration by determining $N\left(r\right)^2$ and $ f\left(r\right)$. To perform the $\tau$ integration, one has to consider the periodicity of $\tau$ given as $\frac{2\pi}{\kappa}$, where $\kappa$ is the surface gravity which in turn is given by,
\begin{equation}
\kappa=\frac{1}{2} \sqrt{N^{\prime} \left(r\right)^2 f\left(r\right)}
\end{equation}
Since we rely on outgoing paths, the integration limit for $\tau$ will be, $0 \leq \tau \leq \frac{\pi}{\kappa}$. Now from Hawking's discovery on temperature of the black holes, we know that temperature of the black is proportional to the surface gravity as,
\begin{equation}
T_{\mbox{bh}}=\frac{\hbar \kappa} {2 \pi}
\end{equation}
Along with these findings, one can directly apply Bohr-Sommerfiled quantization rule to this scenario and that will lead to the area spectrum and eventually the entropy spectrum of the black hole. But the area spectrum as well as the entropy spectrum spacing changes with respect to the change in coordinate transformations. To account for this discrepancy Jiang and Han \cite{jiang} proposed and argued that the closed contour integral $\oint p_i dq_i$ is invariant under coordinate transformations and hence the adiabatic invariant quantity must be of the covariant form, 
\begin{equation}
I=\oint p_i dq_i
\end{equation}
Here the closed contour integral can be considered as a closed path that goes from $q_{i} ^{out}$
(outside the event horizon) to $q_{i} ^{in}$(inside the event horizon). That is,
\begin{equation}
 I=\oint p_i d q_i=\int_{q_{i} ^{in}} ^{q_{i} ^{out}} p_i ^{out} d q_i
 +\int_{q_{i} ^{out}} ^{q_{i} ^{in}} p_i ^{in} d q_i.
 \label{adiabatic covariant action}
\end{equation}
Here $p_i ^{in}$ or $p_i ^{out}$ is the conjugate momentum corresponding to the coordinate $q_i ^{in}$ or $q_{i} ^{out}$, respectively, and $i= 0, 1, 2...$
It is also to be considered that $q^{in}_{1}=r^{in}_{h}(q^{out}_{1}=r^{out}_{h})$ and $q_0 ^{in} \left( q_0 ^{out} \right)=\tau$
where $r_h$ is the horizon radius and $\tau$ is the Euclidean time with a periodicity $\frac{2\pi}{\kappa}$ in which $\kappa$ is the surface gravity. Proceeding the tunneling method using this covariant action, one can arrive at the area spectrum of the black hole. 

This paper is organized as follows. In section \ref{sec2} we use the formalism of Jiang-Han to calculate the entropy spectrum of chargeless BTZ black hole in new massive gravity. Section \ref{sec3} is devoted to the entropy spectrum calculations of  charged black holes in Einstein-Maxwell solution in the context of massive gravity. Finally, section \ref{sec4} contains conclusions of our findings and suggestions for future research.  
\section{Entropy spectrum of chargeless BTZ black hole in new massive gravity}
\label{sec2}
Three dimensional higher derivative gravity model, the New Massive Gravity (NMG) was proposed by Bergshoeff, Hohm and Townsend in 2009 \cite{bht}. 
Its action can be written as an addition of higher curvature term to the usual Einstein-Hilbert action,
\begin{eqnarray}
 S_{\rm NMG} &=&S_{\rm EH}+S_{\rm R },  \\
 \label{NEH} S_{\rm EH}&=& \frac{1}{16\pi G} \int d^3x \sqrt{-g}~
  (R-2\lambda) \\
\label{NFO} S_{\rm R }&=&-\frac{1}{16\pi Gm^2} \int d^3x
            \sqrt{-g}~\Big(R_{\mu\nu}R^{\mu\nu}-\frac{3}{8}R^2\Big),
            \label{NMGAct}
\end{eqnarray}
where $m^2$ is a mass parameter and $G$ is a three dimensional Newton constant. The equation of motion can be obtained as, 
\begin{equation}
 G_{\mu\nu}+\lambda
g_{\mu\nu}-\frac{1}{2m^2}K_{\mu\nu}=0,
\end{equation}
where $G_{\mu\nu}$ is the Einstein tensor given by,
$$G_{\mu\nu}=R_{\mu\nu}-\frac{1}{2}g_{\mu\nu}R,$$ and ,
\begin{eqnarray}
  K_{\mu\nu}&=&2\square R_{\mu\nu}-\frac{1}{2}\nabla_\mu \nabla_\nu R-\frac{\square{}R}{2}g_{\mu\nu}
        +4R_{\mu\rho\nu\sigma}R^{\rho\sigma}\nonumber \\
        &-&\frac{3R}{2}R_{\mu\nu}-R^2_{\rho\sigma}g_{\mu\nu}
         +\frac{3R^2}{8}g_{\mu\nu}.
\end{eqnarray}
The parameters should be chosen in such a way that,  one must end up with a (2+1) BTZ black hole solution \cite{btz1,btz2}. For that, we consider,
\begin{equation}
 m^2=\frac{\Lambda^2}{4(-\lambda+\Lambda)}, \hspace{2cm} \Lambda=- \frac{1}{l^2},
\end{equation}
where $\Lambda$ is the cosmological constant.
From this, one can arrive at the BTZ solution as ,
\begin{equation}
 ds^2_{\rm BTZ}=-f(r)dt^2+\frac{dr^2}{f(r)}+r^2d\phi^2,
\end{equation}
\begin{equation}
 f(r)=-M+\frac{r^2}{\ell^2},
 \label{metric_btz}
\end{equation}
where $M$ is the integration constant corresponding to the mass. 
Now one can calculate the thermodynamic quantities from (\ref{metric_btz}). Then the ADM mass of the black hole can be written as,
\begin{equation}
 M=\frac{r_{+} ^2}{l^2}.
\end{equation}
Hawking temperature is obtained from the relation, $T=\frac{\kappa}{2 \pi}$, as,
\begin{equation}
 T_H= \frac{\sqrt{M}}{2 \pi l}.
\end{equation}
From the ADM mass, entropy of the BTZ black hole can be calculated using either Cardy formula or Wald's formula. We adopt Wald's method to calculate the entropy as,
\begin{equation}
 S=\frac{\pi  r_{+}}{2G} \Big( 1- \frac{1}{2 m^2 l^2} \Big).
 \label{wald_entropy}
\end{equation}
Now we will quantize the entropy of the BTZ black hole  via the adiabatic invariance and Bohr-Sommerfeld quantization rule. According to this rule, 
\begin{equation}
\oint p_i d q_i   =-4i \int _{r_{out}} ^{r_{in}} \int_{0} ^{H} \frac{dH^\prime}{ f\left(r\right)} dr~.
\label{bsrule}
\end{equation}
Using the near horizon approximation, $f(r)$ can be Taylor expanded as,
\begin{equation}
f(r)=f(r) _{r_{+}}  + (r- r_{+}) \frac{d f(r)}{dr} |_{r_{+}} + .......
\end{equation}
Since there exists a pole at $r= r_{+}$ one has to consider a contour integral in such a way that the half loop is going above the pole from right to left, to evaluate the adiabatic invariant integral (\ref{bsrule}).  Using the Cauchy integral theorem, we can arrive at,
\begin{equation}
\oint p_i d q_i   =4\pi \int_{0} ^{H} \frac{dH^\prime}{ \kappa} ~ =2 \hbar \int_{0} ^{H} \frac{dH^\prime}{ T_H} ~ 
\label{bsrule1}
\end{equation}
Now we can write the Smarr formula for the BTZ black hole as,
\begin{equation}
dM=dH=T dS.
\end{equation}
Then (\ref{bsrule1}) becomes,
\begin{equation}
\oint p_i d q_i = 2 \hbar S.
\label{ES}
\end{equation}
The Bohr-Sommerfeld quantization rule is given by,
\begin{equation}
\oint p_i d q_i = 2 \pi n \hbar, ~~~~~~~~~ n=1,2,3,...
\label{BS}
\end{equation}
Comparing (\ref{ES}) and (\ref{BS}), one can arrive at the entropy spectrum as, 
\begin{equation}
S=n \pi
\label{spectra1}
\end{equation}
So the black hole entropy is quantized with a spacing of the entropy spectrum given by,
\begin{equation}
\Delta S= S_{n+1} - S_n = \pi.
\end{equation}
Thus, we see that  entropy spectrum of BTZ black hole in new massive gravity is quantized and
is equally spaced and is independent of the black hole parameters.

\section{Entropy spectrum of charged black hole in Einstein Maxwell solution in the context of massive gravity.}
\label{sec3}
The $3$-dimensional action of Einstein Maxwell massive
gravity with an abelian $U(1)$ gauge field is given by,
\begin{equation}
S=-\frac{1}{16\pi }\int d^{3}x\sqrt{-g}\left[ \mathcal{R}-2\Lambda +L(\mathcal{F})+m^{2}\sum_{i}^{4}c_{i}\mathcal{U}_{i}(g,f)\right] ,
\label{Action}
\end{equation}%
where $\mathcal{R}$, $L(\mathcal{F})$ and $\Lambda$ respectively are , the
scalar curvature, an arbitrary Lagrangian of electrodynamics and the cosmological constant.
The Maxwell invariant is given as $\mathcal{F}=F_{\mu \nu }F^{\mu \nu }$\ , Faraday tensor as $F_{\mu \nu }=\partial _{\mu }A_{\nu
}-\partial _{\nu }A_{\mu }$ and the the gauge potential as $A_{\mu } $.
The $c_{i}$'s are some constants and $\mathcal{U}_{i}$'s are symmetric polynomials of the
eigenvalues of the $d\times d$ matrix $\mathcal{K}_{\nu }^{\mu
}=\sqrt{g^{\mu \alpha }f_{\alpha \nu }}$ which can be written as
\begin{eqnarray}
\mathcal{U}_{1} &=&\left[ \mathcal{K}\right] ,\;\;\;\;\;\mathcal{U}_{2}=%
\left[ \mathcal{K}\right] ^{2}-\left[ \mathcal{K}^{2}\right] ,\;\;\;\;\;%
\mathcal{U}_{3}=\left[ \mathcal{K}\right] ^{3}-3\left[ \mathcal{K}\right] %
\left[ \mathcal{K}^{2}\right] +2\left[ \mathcal{K}^{3}\right] ,  \nonumber \\
&&\mathcal{U}_{4}=\left[ \mathcal{K}\right] ^{4}-6\left[ \mathcal{K}^{2}%
\right] \left[ \mathcal{K}\right] ^{2}+8\left[ \mathcal{K}^{3}\right] \left[
\mathcal{K}\right] +3\left[ \mathcal{K}^{2}\right] ^{2}-6\left[ \mathcal{K}%
^{4}\right] .  \nonumber
\end{eqnarray}
Considering (\ref{Action}) and employing the
variational principle, we can arrive at the field equations as
\begin{equation}
G_{\mu \nu }+\Lambda g_{\mu \nu }-\frac{1}{2}g_{\mu \nu }L(\mathcal{F})-2L_{%
\mathcal{F}}F_{\mu \rho }F_{\nu }^{\rho }+m^{2}\chi _{\mu \nu }=0,
\label{Field equation}
\end{equation}%
\begin{equation}
\partial _{\mu }\left( \sqrt{-g}L_{\mathcal{F}}F^{\mu \nu }\right) =0,
\label{Maxwell equation}
\end{equation}%
where $\chi _{\mu \nu }$ is the massive
term given by,
\begin{eqnarray}
\chi _{\mu \nu } &=&-\frac{c_{1}}{2}\left( \mathcal{U}_{1}g_{\mu \nu }-%
\mathcal{K}_{\mu \nu }\right) -\frac{c_{2}}{2}\left( \mathcal{U}_{2}g_{\mu
\nu }-2\mathcal{U}_{1}\mathcal{K}_{\mu \nu }+2\mathcal{K}_{\mu \nu
}^{2}\right) -\frac{c_{3}}{2}(\mathcal{U}_{3}g_{\mu \nu }-3\mathcal{U}_{2}%
\mathcal{K}_{\mu \nu }+  \nonumber \\
&&6\mathcal{U}_{1}\mathcal{K}_{\mu \nu }^{2}-6\mathcal{K}_{\mu \nu }^{3})-%
\frac{c_{4}}{2}(\mathcal{U}_{4}g_{\mu \nu }-4\mathcal{U}_{3}\mathcal{K}_{\mu
\nu }+12\mathcal{U}_{2}\mathcal{K}_{\mu \nu }^{2}-24\mathcal{U}_{1}\mathcal{K%
}_{\mu \nu }^{3}+24\mathcal{K}_{\mu \nu }^{4}).  \label{massiveTerm}
\end{eqnarray}
In order to obtain the linearly charged three dimensional black hole solutions, consider the metric ansatz as,
\begin{equation}
ds^{2}=-f(r)dt^{2}+f^{-1}(r)dr^{2}+r^{2}d\phi ^{2}.
\label{metric}
\end{equation}
To obtain the exact linearly charged BTZ solutions, we would make the choice of parameters as \cite{cai,hendi},
\begin{equation}
f_{\mu \nu }=diag(0,0,c^{2}h_{ij}),  \label{f11}
\end{equation}
\begin{equation}
\mathcal{U}_{1}=\frac{\left( d-2\right) c}{r},\;\;\;\;\;\mathcal{U}_{2}=%
\mathcal{U}_{3}=\mathcal{U}_{4}=0,  \label{U}
\end{equation}
\begin{equation}
L(\mathcal{F})=-\mathcal{F},
\end{equation}
and as a result of such choice one can arrive at the solution as,
\begin{equation}
f\left( r\right) _{BTZ}=-\Lambda r^{2}-M-2q^{2}\ln \left( \frac{r}{l}%
\right) +m^{2}  r C,  \label{f(r)btz}
\end{equation}
where $M$ and $q$ are related to the mass and charge of the black hole respectively where the $m$ represents the massive term contribution and $C$ is an integration constant

Now one can calculate the thermodynamic quantities using the above metric. 
Hawking temperature can be calculated from the surface gravity on the outer horizon $r_{+}$ as, 
\begin{equation}
T=-\frac{\Lambda r_{+}}{2\pi }-\frac{q^{2}}{2\pi
r_{+}}+\frac{m^{2}C}{4\pi }. 
 \label{temperature}
\end{equation}
For the three dimensional case, entropy of the system takes the form \cite{hawking1,hawking2,hawking3,Hunter1,Hunter2,Hunter3}
\begin{equation}
S=\frac{\pi }{2}r_{+}.  \label{TotalS}
\end{equation}
ADM mass of the black hole can be written as,
\begin{equation}
M=-\Lambda r_{+}^{2}+m^{2}r_{+}C-2q^{2}\ln \left(
\frac{r_{+}}{l}\right). 
\end{equation}
Also the electric potential $\Phi$ can be obtained from the relation $\Phi=\left( \frac{\partial M}{\partial Q} \right)_S$,
\begin{equation}
\Phi=-q \ln \left( \frac{r_+}{l} \right)
\end{equation}
Similar to the above section, we will explore and quantize the entropy of the BTZ black hole via the adiabatic invariance, Bohr-Sommerfeld quantization rule and Cauchy integral theorem.  
Using these, one can easily arrive at (\ref{bsrule1}), given as,
\begin{equation}
\oint p_i d q_i   =4\pi \int_{0} ^{H} \frac{dH^\prime}{ \kappa} ~ =2 \hbar \int_{0} ^{H} \frac{dH^\prime}{ T_H} ~  \nonumber
\end{equation}
Now we can write the Smarr formula for the linearly charged BTZ black hole as,
\begin{equation}
dM=dH=T dS- \Phi dQ
\end{equation}
Then (\ref{bsrule1}) become,
\begin{equation}
\oint p_i d q_i = 2 \hbar S\left[  1+\frac{\Phi}{2Q}  \ln \left( \frac{\mathcal{A}(\mathcal{A} \Lambda - Cm^2 \pi)}{4}\right)     \right]
\label{ES1}
\end{equation}
where $\mathcal{A}$ is the circumference of the 2+1 BTZ black hole system.
Comparing (\ref{ES1}) and (\ref{BS}), one can arrive at the entropy spectrum as, 
\begin{equation}
 S=\frac{n \pi}{2\hbar\left[  1+\frac{\Phi}{2Q}  \ln \left( \frac{\mathcal{A}(\mathcal{A} \Lambda - Cm^2 \pi)}{4}\right)     \right]
}
 \label{spectra2}
\end{equation}
It is interesting to note that the entropy of the BTZ black holes in Einstein Maxwell massive gravity is quantized. From the above relation it is evident that the entropy spectrum depends on the value of the black hole parameters. Considering the  absence of electric charge, one can find that the relation (\ref{spectra2}) reduces to chargeless BTZ black hole case with some numerical differences, discussed in section 2, given by the relation (\ref{spectra1}).

There exists another kind of BTZ solutions in massive gravity, when Einstein-Born-Infeld filed is coupled to the context of massive gravity \cite{hendi}. These are nonlinearly charged BTZ black hole solutions in massive gravity, given as
\begin{equation}
f\left( r\right) =-\Lambda r^{2}-m_{0}+2\beta ^{2}r^{2}\left( 1-\Gamma
\right) +q^{2}\left[ 1-2\ln \left( \frac{r}{2l}\left( 1+\Gamma \right)
\right) \right] +m^{2}Cr, \label{nonlinear}
\end{equation}
where $\beta$ is the nonlinearity parameter which arises from the Born-Infeld Lagrangian, and $\Gamma$ is given by the relation $\Gamma=\sqrt{1+\frac{q^2}{r^2 \beta^2}}$. If one uses the formalism of adiabatic invariance proposed by Majhi and Vagenas to calculate the entropy spectrum of nonlinearly charged  BTZ black hole, it miserably fails in this attempt. So we suggest that the same method should be generalised to the nonlinear solutions too.
\section{Conclusions}
\label{sec4}
In this work, we have investigated the quantization of the entropy of (2+1) dimensional BTZ  black holes in massive gravity models via an adiabatic invariant integral method put forward by Majhi and Vagenas with  modification from Jiang and Han, as well as the Bohr-Sommerfeld quantization rule. We have found that the entropy spectrum is quantized. In the case of chargeless BTZ black hole in New Massive Gravity, the entropy spectrum is quantized, equally spaced and is independent of black hole parameters. On the other hand, in the case of linearly charged BTZ black holes in massive gravity, the entropy spectrum is quantized and the it depends on the black hole parameters such as the circumference of the BTZ black hole.  

\section*{Acknowledgments}
VCK wishes to acknowledge the Associateship of IUCAA, Pune, India.

\end{document}